\documentclass{jnmp}
\usepackage{amsmath}
\usepackage{graphicx}
\JNMPnumberwithin{equation}{section}

\theoremstyle{definition}

\DeclareSymbolFont{AMSb}{U}{msb}{m}{n}
\DeclareMathSymbol{\C}{\mathalpha}{AMSb}{"43}
\DeclareMathSymbol{\R}{\mathalpha}{AMSb}{"52}

\begin{document}

\renewcommand{\evenhead}{Fabian Brau}
\renewcommand{\oddhead}{Lower bounds for the spinless Salpeter equation}

\thispagestyle{empty}

\FirstPageHead{*}{*}{200*}{\pageref{firstpage}--\pageref{lastpage}}{Article}

\copyrightnote{2004}{Fabian Brau}

\Name{Lower bounds for the spinless Salpeter equation}

\label{firstpage}

\Author{Fabian Brau \footnote[1]{FNRS Postdoctoral Researcher}}

\Address{Service de Physique G\'en\'erale et de Physique des Particules \'El\'ementaires, Groupe de Physique Nucl\'eaire Th\'eorique, Universit\'e de Mons-Hainaut, B-7000 Mons, Belgique \\
E-mail: fabian.brau@umh.ac.be\\[10pt]}

\Date{Received Month *, 200*; Revised Month *, 200*; Accepted Month *, 200*}

\begin{abstract}
\noindent
We obtain lower bounds on the ground state energy, in one and three dimensions, for the spinless Salpeter equation (Schr\"odinger equation with a relativistic kinetic energy operator) applicable to potentials for which the attractive parts are in $L^p(\R^n)$ for some $p>n$ ($n=1$ or $3$). An extension to confining potentials, which are not in $L^p(\R^n)$, is also presented.
\end{abstract}

\section{Introduction}
\label{sec1}

The spinless Salpeter equation is a simple relativistic version of the Schr\"odinger equation which can be obtained, with some approximations, from the Bethe-Salpeter equation \cite{salp51,grei94}. A covariant description of bound states of two particles is achieved with the Bethe-Salpeter equation \cite{salp51}. This equation reduces to the spinless Salpeter equation \cite{grei94} when the following approximations are performed:  
\begin{itemize}
\item elimination of any dependences on timelike variables (which leads to the Salpeter equation \cite{salp52}).
\item any references to the spin degrees of freedom of particles are neglected as well as negative energy solutions.
\end{itemize}
 
The spinless Salpeter equation takes the form ($\hbar=c=1$)
\begin{equation}
\label{eq1}
\left[\alpha\sqrt{{\bf p}^2+m^2}+V({\bf r})\right]\, \Psi({\bf r})=M\, \Psi({\bf r}),
\end{equation}
where $m$ is the mass of the particle and $M$ is the mass of the eigenstate ($M=\alpha m+E$, $E$ is the binding energy). The parameter $\alpha$ equals $1$ for a one-particle problem and equals $2$ for a two-particle (identical) problem (in the center of mass frame). We restrict our attention to interactions which are introduced in the free equation through the substitution $M \rightarrow M-V({\bf r})$, where $V({\bf r})$ is the time component of a relativistic four-vector. The interaction could also, in principle, be introduced through the substitution ${\bf p} \rightarrow {\bf p}-{\bf A}({\bf r})$, where ${\bf A}({\bf r})$ is the spatial component of a relativistic four-vector. However we do not consider this kind of potentials since the derivation of the spinless Salpeter equation from the Bethe-Salpeter equation leads to ${\bf A}({\bf r})=0$. Equation (\ref{eq1}) is generally used when kinetic relativistic effects cannot be neglected and when the particles under consideration are bosons or when the spin of the particles is neglected or is only taken into account via spin-dependent interactions. Despite its apparent complexity, this equation is often preferred to the Klein-Gordon equation. The equation (\ref{eq1}) appears, for example, in mesons and baryons spectroscopy in the context of potential models\footnote{For a review of several aspects of the ``semirelativistic" description of bound states with the spinless Salpeter equation see: W. Lucha and F. F. Sch\"oberl, Semirelativistic treatment of bound states, \textit{Int. J. Mod. Phys. A} \textbf{14} (1999), 2309--2334 and references therein.} (see for example \cite{go85,se97,br98,gl98,br02}). 

Conversely to the Schr\"odinger equation, for which a fairly large number of results, giving limits on the number of bound states or limits on the values of energy levels, can be found in the literature (see for example \cite{ba52,sc61,ca65a,ch68,ha69,ba76,gl76,ma77,ma78,ba78,ba79,li80,fe86,ch95a,oc01,br03a,br03b,br03c,br04a,br04b}), there are few general rigorous results concerning the spinless Salpeter equation due to the pseudo-differential nature of the kinetic energy operator. Most of these results have been obtained for a Coulomb potential (for example, upper and lower bounds on energy levels) \cite{he77,ca84,ha84,ma89,ra94}. Recently upper and lower limits on energy levels have been obtained for some other particular interactions \cite{ha01a,ha01b,ha02,ha03}. Some more general results also exist. An upper bound on the total number of bound states in three dimensions and an upper limit on the number of $\ell$-wave bound states have been obtained in Refs. \cite{daub83,br03d}. An upper limit on the critical value, $g_{\rm{c}}^{(\ell)}$, of the coupling constant (strength), $g$, of the central potential, $V(r)=g\, v(r)$, for which a first $\ell$-wave bound state appears has been derived in Ref. \cite{br04c}. 

In this article we obtain lower bounds on the ground state energy, in one and three dimensions, for the spinless Salpeter equation applicable to potentials for which the attractive parts are in $L^p(\R^n)$ for some $p>n$ ($n=1$ or $3$). These lower bounds yield lower limits on a second kind of critical value, $\tilde{g}_{\rm{c}}$, of the coupling constant, $g$, of the potential, $V({\bf r})=g\, v({\bf r})$, for which the mass of the eigenstate $M$ is vanishing, entailing that for any value of $g$ greater than $\tilde{g}_{\rm{c}}$ the system has an unphysical negative mass.

\section{General formulas}
\label{sec2}

In this Section, we derive some general formulas that we apply in Sections \ref{sec3} and \ref{sec4} to obtain the lower bounds in one and three dimensions.

We begin by recalling the expression of the sharp Young's inequality obtained in 1975 by Beckner and independently by Brascamp and Lieb \cite{be75,br76}
\begin{equation}
\label{eq2}
\left|\int d^n{\bf x}\,d^n{\bf y}\, f({\bf x})\, g({\bf x}-{\bf y})\, h({\bf y})\right|\leq \left[C_p C_q C_r\right]^n\, ||f||_p\, ||g||_q\, ||h||_r,
\end{equation}
where $p^{-1}+q^{-1}+r^{-1}=2$ with $p$, $q$, $r\geq 1$ and
\begin{equation}
\label{eq3}
||f||_p=\left[\int d^n{\bf x}\, |f({\bf x})|^p \right]^{1/p}.
\end{equation}
The constants appearing in the inequality (\ref{eq2}) are best possible and take the form
\begin{equation}
\label{eq4}
(C_p)^2=p^{1/p}\, p'^{-1/p'},
\end{equation}
where $p'^{-1}=1-p^{-1}$.
  
Now we write the spinless Salpeter equation (\ref{eq1}) as an integral equation
\begin{equation}
\label{eq5}
\Psi({\bf x})=\int d^n{\bf y}\, G^{(n)}(\Delta)\, [M-V({\bf y})]\, \Psi({\bf y}),
\end{equation}
where ${\bf \Delta}={\bf x}-{\bf y}$, $\Delta=|{\bf \Delta}|$ and $G^{(n)}(\Delta)$ is the Green's function of the kinetic energy operator, $\alpha\,[{\bf p}^2+m^2]^{1/2}$, in $n$ dimensions. For $n=3$ the expression of the Green's function is known \cite{brau98} (see also Section \ref{sec3}). For $n=1$ the Green's function is obtained in Section \ref{sec4}.

The Young's inequality (\ref{eq2}) can be used together with the expression (\ref{eq5}) of the spinless Salpeter equation to obtain
\begin{equation}
\label{eq6}
||\Psi||_2^2\leq |M| \left[C_{\tilde{p}} C_{\tilde{q}} C_{\tilde{r}}\right]^n\, ||\Psi||_{\tilde{p}}\, ||G^{(n)}||_{\tilde{q}}\, ||\Psi||_{\tilde{r}}+ \left[C_p C_q C_r\right]^n\, ||V^- \Psi||_p\, ||G^{(n)}||_q\, ||\Psi||_r,
\end{equation}
where $V^-({\bf x})=\max(0,-V({\bf x}))$ is the attractive (negative) part of the potential. In order to obtain a formula without any reference to the wave function, we choose ${\tilde{p}}={\tilde{r}}=2$, ${\tilde{q}}=1$ and $r=2$. With the help of H\"older's inequality applied to $||V^- \Psi||_p$, it is possible to write
\begin{equation}
\label{eq7}
||\Psi||_2^2\leq |M|\, ||G^{(n)}||_1 \, ||\Psi||_2^2 + [C_q C_{\frac{2q}{3q-2}}]^n\, ||V^-||_{\frac{q}{q-1}}\, ||G^{(n)}||_q\, ||\Psi||_2^2
\end{equation}
with $q\geq 1$. We obviously suppose for now that $||G^{(n)}||_q$ exists for some $1\leq q< q^{(n)}_0$. This question is discussed in Sections \ref{sec3} and \ref{sec4}. The quantity $||V^-||_{\frac{q}{q-1}}$ exists provided the attractive part of the potential is in $L^p(\R^n)$ for some $p>q^{(n)}_0/(q^{(n)}_0-1)$. 

From (\ref{eq7}) we finally obtain the following lower limits
\begin{subequations}
\label{eq8}
\begin{equation}
\label{eq8a}
|M|\geq ||G^{(n)}||_1^{-1}\left[1-{\cal C}_q^n\, ||V^-||_{\frac{q}{q-1}}\, ||G^{(n)}||_q\right],
\end{equation}
where the constant takes the form
\begin{equation}
\label{eq8b}
{\cal C}^2_q=[C_q C_{\frac{2q}{3q-2}}]^2=q^{1/q}\, \left(\frac{q-1}{q}\right)^{(q-1)/q}\, \left(\frac{2q}{3q-2}\right)^{(3q-2)/2q}\, \left(\frac{2-q}{2q}\right)^{(2-q)/2q}.
\end{equation}
\end{subequations}
Physically $M$ is required to be nonnegative and the quantity $|M|$ could be replaced by the mass itself, $M$, provided that we restrict the right-hand side of (\ref{eq8a}) to be nonnegative.
 
Equation (\ref{eq8}) is the main result of this article. In the next Sections we apply this relation to obtain lower bounds in one and three dimensions.

\section{Lower bound in three dimensions}
\label{sec3}

In order to apply the relation (\ref{eq8}) in the specific case of a three-dimensional space we need the relevant Green's function. The computation has already been done in \cite{brau98}. The Green's function is obtained by computing the following integral
\begin{equation}
\label{eq10}
G^{(3)}(\Delta)=\frac{1}{(2\pi)^3}\int d{\bf p}\, \frac{\exp(-i\, {\bf p}\cdot {\bf \Delta})}
{\alpha\sqrt{p^2+m^2}}.
\end{equation}
We find the following expression
\begin{equation}
\label{eq11}
G^{(3)}(\Delta)=\frac{m}{2\alpha \pi^2 \Delta}K_1(m\Delta),
\end{equation}
where $K_{\nu}(x)$ is a modified Bessel's function (see for example \cite[p. 374]{abra70}).
With the expression (\ref{eq11}) for the Green's function we can compute the needed norms. We find that
\begin{equation}
\label{eq12}
||G^{(3)}||_1=\frac{1}{\alpha m}
\end{equation}
and
\begin{eqnarray}
\label{eq13}
||G^{(3)}||_q&=&m^{2-3/q}\, \frac{(4\pi)^{1/q}}{2\alpha \pi^2}\, \left[\int_0^{\infty}dx\, x^{2-q}\, K_1(x)^q\right]^{1/q}\nonumber \\
     &\equiv& m^{2-3/q}\, \frac{\tilde{{\cal C}}_q}{\alpha}.
\end{eqnarray}
These relations, (\ref{eq12}) and (\ref{eq13}), imply that we cannot consider the ultrarelativistic regime ($m=0$) in order to have finite values for the various norms of the Green's function. Moreover, since the Bessel's function $K_1(x)$ behaves like $x^{-1}$ for $x\approx 0$ ($\lim_{x\rightarrow 0} xK_1(x)=1$), we have the following restriction on $q$ in order to have a finite value of the constant $\tilde{{\cal C}}_q$
\begin{equation}
\label{eq14}
1\leq q<q^{(3)}_0=\frac{3}{2}.
\end{equation}

The lower bound on the mass of the ground state is (we suppose $M$ to be nonnegative)
\begin{equation}
\label{eq15}
M\geq \alpha\, m -{\cal C}_q^3\, \tilde{{\cal C}}_q\, m^{3-3/q}\, ||V^-||_{\frac{q}{q-1}}.
\end{equation}
This lower limit is nontrivial only if $||V^-||_{\frac{q}{q-1}}$ exists. The restriction (\ref{eq14}) implies that the negative part of the potential must be in $L^p(\R^3)$ for some $p>3$. Note that the limit $q\downarrow 1$ of the inequality (\ref{eq15}) gives $M\geq \alpha m-||V^-||_{\infty}$ which is trivially true.
 
The inequality (\ref{eq15}) obviously yields the following lower bound on the ground state binding energy
\begin{equation}
\label{eq16}
E\geq -{\cal C}_q^3\, \tilde{{\cal C}}_q\, m^{3-3/q}\, ||V^-||_{\frac{q}{q-1}}.
\end{equation}
This lower limit implies that the binding energy does not increase, in modulus, faster than linearly with the mass of the particle and with the strength of the potential (for the class of potentials we consider).

We propose now to test the lower bound (\ref{eq15}) with some typical potentials. For simplicity we consider only central potentials. We use three potentials for these tests: an exponential potential
\begin{equation}
\label{eq16a}
V(r)=-gR^{-1} \exp(-r/R);
\end{equation}
an exponential multiplied by a power
\begin{equation}
\label{eq16b}
V(r)=-gR^{-2}\, r\exp(-r/R);
\end{equation}
and a singular potential
\begin{equation}
\label{eq16c}
V(r)=-g (rR)^{-1/2}\exp(-r/R).
\end{equation}

Note that the lower bound (\ref{eq15}) does not apply to the Yukawa potential, for example, since it behaves as $r^{-1}$ near the origin and thus the negative part of this potential is not in $L^p(\R^3)$ for any $p>3$.

The mass of the ground state depends on two dimensionless parameters, $g$ and $\beta=mR$. It is thus rather cumbersome to compare the lower bound (\ref{eq15}) to the exact result. We propose an indirect test of this lower bound by computing a lower bound on the critical value, $\tilde{g}_{\rm{c}}$, of the potential, $V(r)=g\, v(r)$, for which the mass $M$ of the ground state is vanishing, entailing that for any value of $g$ greater than $\tilde{g}_{\rm{c}}$ the system has an unphysical negative mass. From (\ref{eq15}) we obtain the lower limit
\begin{equation}
\label{eq16d}
\tilde{g}_{\rm{c}}\geq \alpha \left[{\cal C}_q^3\, \tilde{{\cal C}}_q\, m^{2-3/q}\, ||v^-||_{\frac{q}{q-1}}\right]^{-1}.
\end{equation}
This inequality depends only on one parameter, $\beta$, and can then be tested more easily.

\begin{figure}[t]
\begin{center}
\includegraphics*[width=12cm]{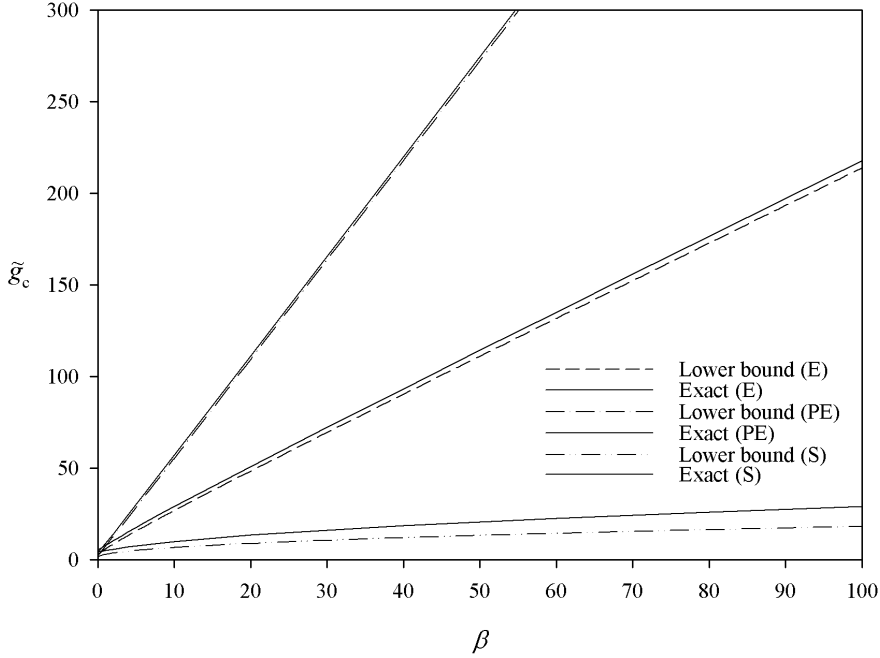}
\end{center}
\caption{Comparison between the exact values of the critical coupling constant, $\tilde{g}_{\rm{c}}$, obtained numerically and the lower limit (\ref{eq16d}) as a function of $\beta=mR$ for the three potentials (\ref{eq16a}), (\ref{eq16b}) and (\ref{eq16c}), denoted respectively E, PE and S.}
\label{fig1}
\end{figure}

In Figure \ref{fig1}, we present a comparison between the exact values of the critical coupling constant, $\tilde{g}_{\rm{c}}$, obtained numerically, and the lower limit (\ref{eq16d}) as a function of $\beta=mR$ for the three potentials (\ref{eq16a})-(\ref{eq16c}). Note that for these tests we choose to solve the spinless Salpeter equation with two identical particles, $\alpha=2$, only for numerical convenience (not to modify numerical codes). Clearly the lower bound (\ref{eq16d}) is quite cogent and follow closely the exact behavior for all values of $\beta$. The accuracy is, however, less good for the singular potential.

\section{Extension to confining potentials}
\label{sec3b}

In this Section we consider confining potentials, such as the harmonic potential, for which the norm $||V^-||_{\frac{q}{q-1}}$ does not exist. Then the lower bound (\ref{eq15}) does not apply without suitable modifications. This kind of potential is used, for example, in mesons and baryons spectroscopy, in the framework of potential models, to describe quark confinement together with the spinless Salpeter equation \cite{go85,se97,br98,gl98,br02}. 

To obtain a lower bound for this class of potentials we consider the following truncated potential $\tilde{V}(r)$ defined as
\begin{eqnarray}
\label{eq16e}
\tilde{V}(r)&=&V(r) \quad \text{for}\quad V(r)\leq C \nonumber \\
  &=& C \quad \text{for}\quad V(r)> C.
\end{eqnarray}
Since $\tilde{V}(r)$ is more attractive than $V(r)$, a lower bound on the ground state energy of $\tilde{V}(r)$ is also a lower bound on the ground state energy of $V(r)$. To be able to compute the relevant norm of the interaction we consider the potential $\bar{V}(r)\equiv \tilde{V}(r)-C$. Now the lower bound (\ref{eq15}) applies to $\bar{V}(r)$ and we have
\begin{equation}
\label{eq16g}
|M_{\bar{V}}|\geq \alpha\, m \left[1-{\cal C}_q^3\, \tilde{{\cal C}}_q\, m^{2-3/q}\, ||\bar{V}^-||_{\frac{q}{q-1}}\right]\equiv \bar{M},
\end{equation}
where $M_{\bar{V}}$ denotes the mass of the ground state of the potential $\bar{V}(r)$ and where we consider, for a moment, the general form of the lower bound with the modulus of $M_{\bar{V}}$. The relation between the potentials, $\tilde{V}(r)$ and $\bar{V}(r)$, clearly implies that
\begin{equation}
\label{eq16h}
M_{\tilde{V}}= M_{\bar{V}}+C.
\end{equation}
However we do not have a relation between $M_{\tilde{V}}$ and $|M_{\bar{V}}|$ and we must restrict the lower bound (\ref{eq16g}) to nonnegative values of $M_{\bar{V}}$, which is secured if $\bar{M}\geq 0$.
This restriction implies a limitation on the possible values of the constant $C$ which cannot be too large. We can then write
\begin{equation}
\label{eq16i}
M\geq M_{\tilde{V}}=M_{\bar{V}}+C\geq \bar{M}+C.
\end{equation}

We have two parameters, $q$ and $C$, that can be used to maximize the right-hand side of (\ref{eq16i}) with the restrictions $1\leq q<3/2$ and $\bar{M}\geq 0$. Actually the best optimization is obtained by searching the value $q^*$ of $q$ which yields the greatest value $C^*$ of $C$ such that $\bar{M}=0$. In other words we search for the value $q^*$ of $q$ which yields the greatest value $C^*$ of $C$ such that
\begin{equation}
\label{eq16j}
{\cal C}_q^3\, \tilde{{\cal C}}_q\, m^{2-3/q}\, ||\bar{V}^-||_{\frac{q}{q-1}}=1.
\end{equation}
This procedure yields the lower bound,
\begin{equation}
\label{eq16k}
M\geq C^*,
\end{equation}
once $C^*$ has been found. Note that the limit $q\downarrow 1$ of the inequality (\ref{eq16k}) yields $M\geq \alpha m+\min(V(r))$ which is trivially true.

\begin{figure}[t]
\begin{center}
\includegraphics*[width=12cm]{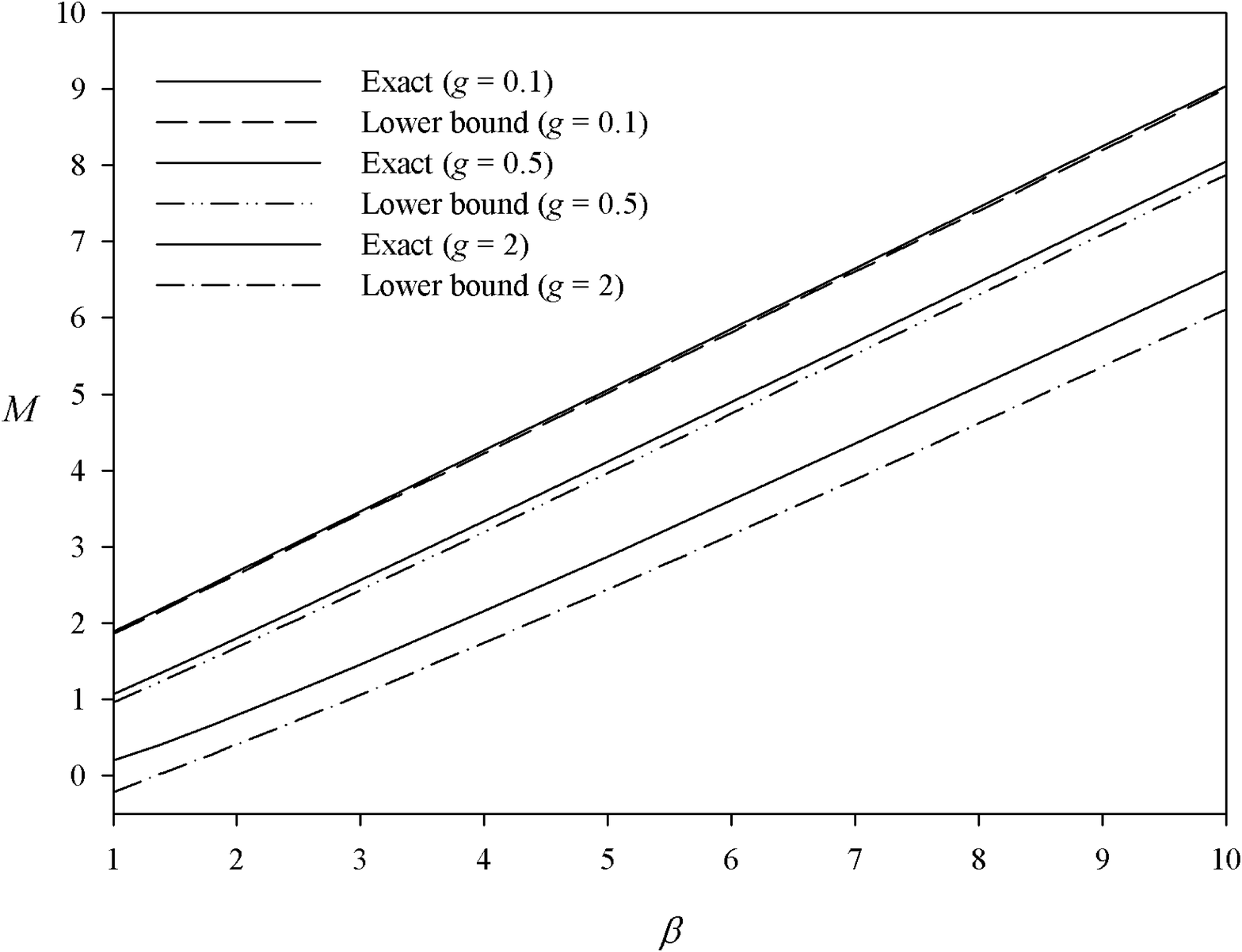}
\end{center}
\caption{Comparison between the exact values of the mass $M$ (in GeV) of the ground state of the logarithmic potential (\protect\ref{eq16l}) and the lower bound (\ref{eq16k}) as a function of $\beta=mR$ for three values of $g$. 1 GeV has been added to the masses for $g=0.1$ and 1 GeV has been subtracted from the masses for $g=2$ (to obtain a clearer presentation).}
\label{fig2}
\end{figure}

We propose to test this lower bound with a central logarithmic potential
\begin{equation}
\label{eq16l}
V(r)=gR^{-1}\, \ln(r/R).
\end{equation}
Quigg and Rosner introduced this potential in 1977 to describe the spectrum of heavy mesons \cite{qui77}. The values of the parameters they found was $R\approx 2.5$ GeV$^{-1}$ and $g\approx 1.8$. Note that these values of the parameters were obtained with the Schr\"odinger equation and not with the spinless Salpeter equation.

In Figure \ref{fig2} we present a comparison between the exact values of the mass of the ground state of the logarithmic potential (\protect\ref{eq16l}) and the lower bound (\ref{eq16k}) as a function of $\beta=mR$ for three values of $g=0.1$, $0.5$, $2$. More precisely we fix $R$ to $2.5$ GeV$^{-1}$ and we let $m$ vary between $0.4$ GeV and $4$ GeV, which covers approximatively the various constituent (effective) masses of the quarks used in potential models (from the $u$ and $d$ quarks to the $b$ quark). We here also choose to solve numerically the spinless Salpeter equation with two identical particles, $\alpha=2$. Again the accuracy of the lower bound is quite satisfactory, especially for small values of $g$.

\section{Lower bound in one dimension}
\label{sec4}

To conclude this article we derive the lower bound applicable to one-dimensional space. Again, in order to apply the relation (\ref{eq8}), we need the relevant Green's function. It is obtained by computing the following integral
\begin{equation}
\label{eq17}
G^{(1)}(\Delta)=\frac{1}{2\pi}\int_{-\infty}^{+\infty} dp\, \frac{\exp(-i\, p \Delta)}{\alpha\sqrt{p^2+m^2}}.
\end{equation}
We find the following expression for the Green's function
\begin{equation}
\label{eq18}
G^{(1)}(\Delta)=\frac{1}{\alpha \pi}K_0(m\Delta),
\end{equation}
where $K_{\nu}(x)$ is again a modified Bessel's function. With the expression (\ref{eq18}) for the Green's function, we obtain
\begin{equation}
\label{eq19}
||G^{(1)}||_1=\frac{1}{\alpha m}
\end{equation}
and
\begin{eqnarray}
\label{eq20}
||G^{(1)}||_q&=&m^{-1/q}\, \frac{2^{1/q}}{\alpha \pi}\, \left[\int_0^{\infty}dx\,  K_0(x)^q\right]^{1/q}\nonumber \\
     &\equiv& m^{-1/q}\, \frac{\bar{{\cal C}}_q}{\alpha}.
\end{eqnarray}
These relations, (\ref{eq19}) and (\ref{eq20}), imply that we cannot consider the ultrarelativistic regime ($m=0$) in order to have finite values for the various norms of the Green's function. The Bessel's function $K_0(x)$ is characterized by a logarithmic singularity for $x\approx 0$ and the constant $\bar{{\cal C}}_q$ is well defined for $q\geq 1$ ($q_0^{(1)}=\infty$).

The lower bound on the ground state energy is, where $M$ is supposed to be nonnegative,
\begin{equation}
\label{eq21}
M\geq \alpha\, m -{\cal C}_q\, \bar{{\cal C}}_q\, m^{1-1/q}\, ||V^-||_{\frac{q}{q-1}}.
\end{equation}
This lower limit is nontrivial only if $||V^-||_{\frac{q}{q-1}}$ exists. Thus the negative part of the potential must be in $L^p(\R)$ for some $p>1$. 

\section*{Acknowledgments}

We thank Norbert Euler, Managing Editor of the Journal of Nonlinear Mathematical Physics, for his kind invitation to write this article on the occasion of the seventieth birthday of Francesco Calogero.

\label{lastpage}

\end{document}